\documentclass[journal]{IEEEtran}
\usepackage{graphicx} 
\usepackage{multicol}
\usepackage{fancyhdr}
\usepackage{lipsum}
\usepackage{etoolbox}
\usepackage{color}
\usepackage{amsmath}
\usepackage[colorlinks=true, allcolors=blue]{hyperref}
\usepackage{cite,epsfig,amssymb,amsmath,epstopdf,algorithm,algorithmic,subfigure,bbding,caption}
\newif\iffirstpage
\firstpagetrue

\title{Responsible Federated Learning in Smart Transportation: Outlooks and Challenges}
\author{Xiaowen Huang, Tao Huang$^{\dag}$, Shushi Gu, Shuguang Zhao, Guanglin Zhang$^{\dag}$}

\begin{document}

\maketitle

\begin{abstract}

Integrating artificial intelligence (AI) and federated learning (FL) in smart transportation has raised critical issues regarding their responsible use. Ensuring responsible AI is paramount for the stability and sustainability of intelligent transportation systems. Despite its importance, research on the responsible application of AI and FL in this domain remains nascent, with a paucity of in-depth investigations into their confluence. Our study analyzes the roles of FL in smart transportation, as well as the promoting effect of responsible AI on distributed smart transportation. Lastly, we discuss the challenges of developing and implementing responsible FL in smart transportation and propose potential solutions. By integrating responsible AI and federated learning, intelligent transportation systems are expected to achieve a higher degree of intelligence, personalization, safety, and transparency.
\end{abstract}

\iffirstpage
 \footnotetext{Manuscript accepted by IEEE Internet of Things Magazine 6 Feb 2024.  This work was supported in part by the National Natural Science Foundation of China under Grant 62072096, in part by the Program of Shanghai Academic/Technology Research Leader 23XD1420100, and in part by the Program for Professor of Special Appointment (Eastern Scholar) at Shanghai Institutions of Higher Learning. (\textit{Corresponding author: Tao Huang, Guanglin Zhang.})
 
 Xiaowen Huang, Shuguang Zhao, and Guanglin Zhang are with the College of Information Science and Technology, Donghua University, Shanghai, China (email: huangxiaowen@mail.dhu.edu.cn; sgzhao@dhu.edu.cn; glzhang@dhu.edu.cn).

 Tao Huang is with the James Cook University, Cairns, Australia (email: tao.huang1@jcu.edu.au).
 
 Shushi Gu is with the School of Electronic and Information Engineering in Harbin Institute of Technology (Shenzhen), Shenzhen, China, and the researcher with Guangdong Provincial Key Laboratory of Aerospace Communication and Networking Technology, Shenzhen, China (email: gushushi@hit.edu.cn).

 \copyright 2024 IEEE. Personal use of this material is permitted. Permission from IEEE must be obtained for all other uses, in any current or future media, including reprinting/republishing this material for advertising or promotional purposes, creating new collective works, for resale or redistribution to servers or lists, or reuse of any copyrighted component of this work in other works.}
  \global\firstpagefalse 
\fi

\section{Introduction}

In the 2016 paper ``Concrete Problems in AI Safety," Dario Amodei raised concerns about the ethical implications of artificial intelligence (AI) \cite{paper_001}. These concerns were later reflected in Google's 2018 AI Principles \cite{paper_002}, which aimed to promote ethical practices in the field of AI. However, as AI becomes more widely used, its ethical, legal, and social implications have become a more pressing concern for a wide range of stakeholders, including technicians, policymakers, ethicists, and the general public.

The advent of responsible AI responds to these issues, striving to align AI's development and deployment with ethical and legal tenets. This paradigm extends beyond technicalities to encompass ethical considerations and the human element in AI's decision-making, with an overarching intent to enhance human welfare. Microsoft's 2022 Responsible AI Standard \cite{paper_003} epitomizes this ethos, offering a framework rooted in six fundamental tenets: fairness, reliability, security, privacy, inclusiveness, transparency, and accountability, guiding AI's evolution towards ethical congruence.

Achieving responsible AI requires collaborative efforts of various stakeholders. Ethicists play a crucial role in formulating ethical principles and guidelines to guide the development and use of AI. Policymakers must establish appropriate regulations and policies that govern the use of AI technologies. Technicians should develop AI systems that are transparent, fair, and explainable. Most importantly, the public must be informed and involved in the decision-making processes surrounding AI technologies to ensure that the technology aligns with community interests and dignity.

\begin{figure}[t]
\centering
\includegraphics[scale=0.5]{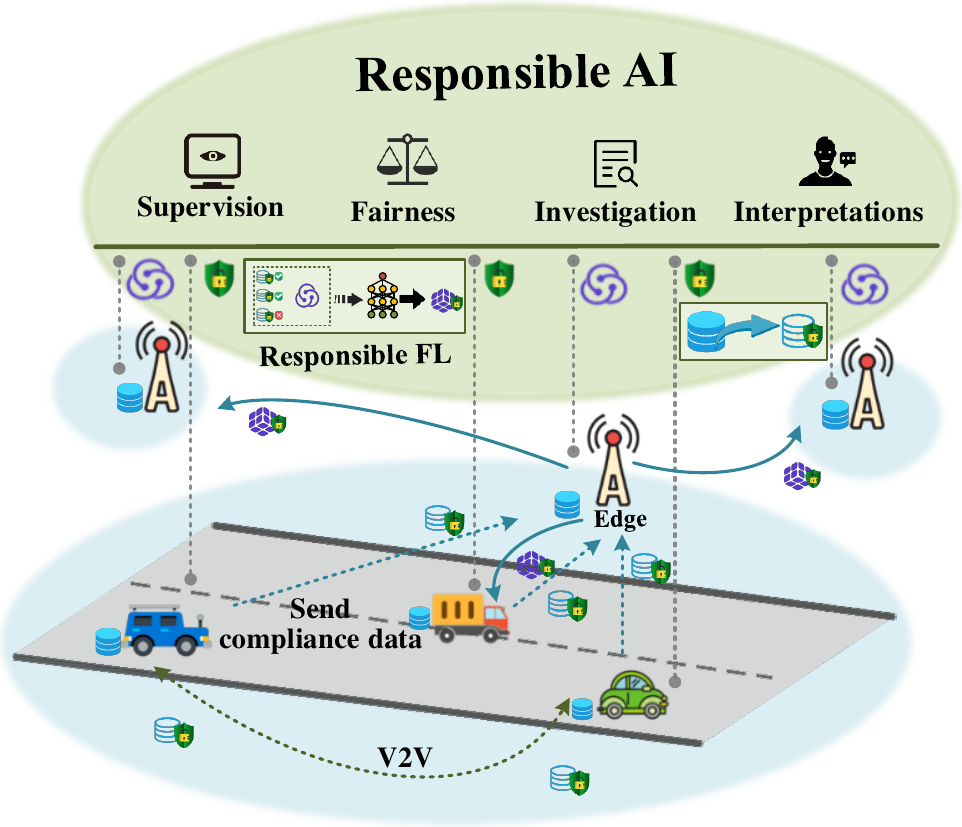} 
\caption{The examples of responsible AI and FL in smart transportation.}
\label{fig1}
\end{figure}

Federated learning (FL) has emerged as a trusted AI paradigm that enables secure and efficient collaborative computation among multiple participants \cite{paper_004}. This approach allows multiple participants to train AI models collectively without sharing their raw data, thereby safeguarding data privacy and confidentiality. Each participant in FL retains ownership of its dataset and conducts local model training. The results of the local training are then uploaded to a central server, which performs model aggregation and updating. FL enables vehicles to train models without compromising data privacy while improving the accuracy and generalization of smart transportation models.

There is a growing need for responsible AI and FL in smart transportation. However, there is limited research on their role in this context \cite{paper_005,paper_006,paper_007,paper_008}, as shown in Table 1.  Therefore, this paper explores responsible FL in smart transportation and identifies challenges and opportunities for future research efforts. For instance, as illustrated in Fig. \ref{fig1}, vehicles transmit parameter data to edge nodes, which undergo a selection process to identify trustworthy information for subsequent analysis and aggregation. In the context of driver instructions, vehicles are expected to provide comprehensive explanations for recommended actions. The transportation system follows a well-defined responsibility framework, allowing for the identification of error-causing processes in the event of failures.

\begin{figure}[t]
\centering
\includegraphics[scale=0.5]{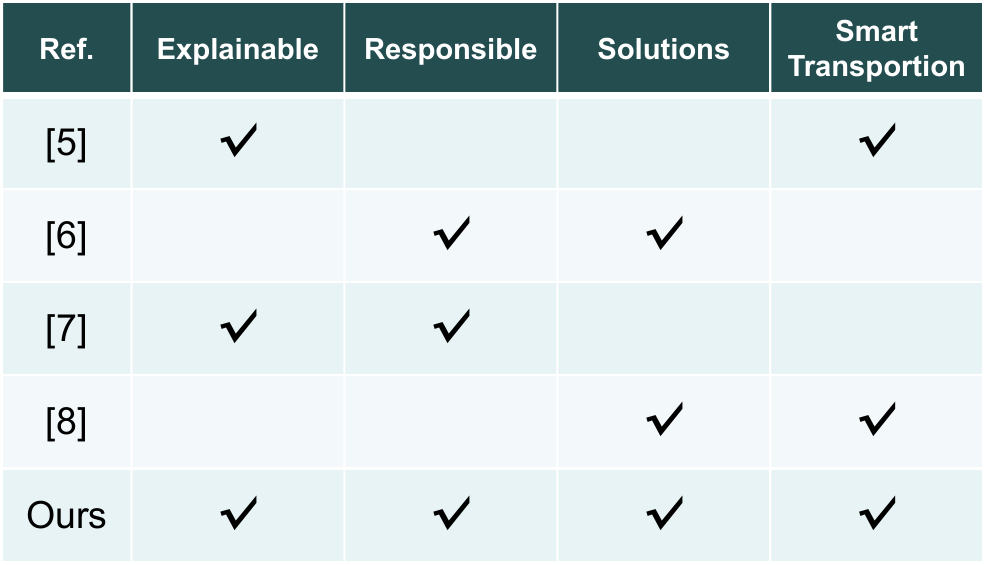} 
\caption*{Table 1. Comparison of existing literature and our work.}
\label{table1}
\end{figure}

In this paper, we elucidate the advantages of FL within the smart transportation sector. Our analysis extends to the incorporation of a responsible AI framework tailored to this domain, highlighting the pivotal elements of responsible FL deployment in smart transportation systems. Additionally, we rigorously examine the inherent challenges posed by the integration and enactment of responsible FL in this context, proposing a suite of plausible resolutions. This contribution not only synthesizes current knowledge but also charts a course for future research and application in responsible FL for smart transportation. 

Introducing responsible FL in smart transportation can enhance system design, technology development, and theoretical optimization, promoting real-time efficiency, sustainability, and compliance with improved laws and regulations. This integration fosters a comprehensive and responsible approach, optimizing transportation management and advancing the industry's intelligent and green development. Predictably, responsible FL has profound and diverse implications in smart transportation. For instance, it can enhance the accuracy of traffic accident warning systems, reducing public property and human losses. It can also predict and adjust the duration of traffic signal lights in real-time, alleviating traffic pressure. Explaining the principles and methods of decision-making processes can enhance user trust and acceptability, while personalized services significantly improve users' fine-grained experience. Next, in the ongoing operation of the system, it can quickly respond to changes in traffic conditions and reduce operating costs because of its distributed model training. Lastly, it provides possibilities for collaboration in different fields, such as transportation, energy, and environmental protection, leading to mutually beneficial outcomes. Relevant regulations and policies also need to keep pace with the times, providing support for emergency response.

\section{FL Services for Smart Transportation}

\begin{figure}[t]
\centering
\includegraphics[scale=1]{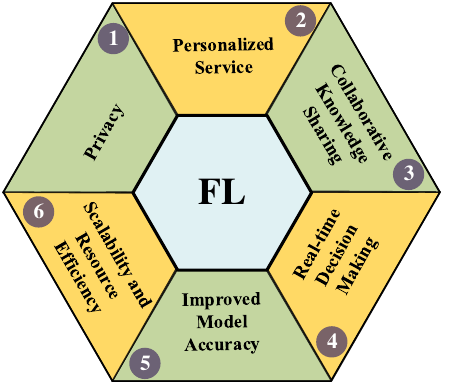} 
\caption{Advantages of FL for smart transportation.}
\label{fig2}
\end{figure}

FL technology enhances the fundamental services of the Industrial Internet of Things (IIoT) and offers more potential services based on characteristics such as privacy preservation, distribution, and collaboration. In this section, we briefly discuss the advantages of FL for smart transportation, as depicted in Fig. \ref{fig2}.

\subsection{Privacy}

Smart transportation systems rely on the analysis of sensitive data from various sources, such as cameras, speedometers, etc. However, sharing raw data poses a risk to private information such as location or vehicle sensor data. To solve this problem, FL enables machine learning models to be trained directly on local vehicles or local roadside units (RSUs), which ensures data decentralization and privacy. Furthermore, combined with blockchain, trusted technologies and more, FL can further improve data security \cite{paper_009}. By implementing FL, smart transportation systems can mitigate the risk of data leakage, misuse, and unauthorized access, thereby enhancing user trust and stakeholder confidence.

\subsection{Personalized Service}

Smart transportation offers personalized travel mode and route recommendations, optimizing convenience and efficiency. It also provides tailored suggestions for parking and nearby services, catering to individual preferences. FL addresses the challenge of balancing personalized service provision and data protection. By utilizing FL, personalized models capture individual preferences and adapt to real-time factors like traffic conditions and user location, enabling personalized recommendations for parking, routes, transportation modes, and nearby services \cite{paper_010}. FL facilitates continuous learning and improvement of personalized models while giving users control over their data and customization options. This ensures that personalized services align with users' specific requirements and privacy preferences.

\subsection{Collaborative Knowledge Sharing}

For effective smart transportation systems, harmonizing devices from various producers, ensuring stable vehicle-to-vehicle (V2V) connectivity, and adeptly handling diverse data types are essential. Critical to this infrastructure's integrity are stringent cybersecurity and privacy safeguards. FL enhances model accuracy via aggregation while maintaining privacy, circumventing the need to exchange raw data \cite{paper_011}. FL thus fosters cooperative engagement among all participants in the smart transportation network, such as automakers, infrastructure entities, and service providers.

\subsection{Real-time Decision Making}

Smart transportation involves managing various aspects such as traffic signal control, vehicle scheduling, parking management, public transportation, and emergency management. Real-time decision-making is required to address the dynamic traffic changes in these areas. It helps optimize traffic flow, reduces congestion, and enables timely accident response \cite{paper_012}. FL supports real-time decision-making by utilizing distributed real-time data processing, predictive modelling, collaborative decision-making, on-device intelligence, and adaptive context awareness. FL empowers local and edge devices like vehicles and RSU to make intelligent real-time decisions. FL training models directly on these devices enable collaborative decision-making through knowledge aggregation and on-device intelligence. This reduces the reliance on centralized servers and enhances system responsiveness, minimizing continuous data transmission and processing.

\subsection{Improved Model Accuracy}

Improved transportation systems require more accurate predictions of traffic flow, speeds, and accident risks, as well as smarter vehicle scheduling and route planning, enhanced safety and security, and optimized resource allocation. To achieve this, FL enables decentralized data collaboration, enhances data diversity, and supports real-time model updates. FL offers benefits such as increased model accuracy and reliability in transportation applications including traffic prediction, route optimization, anomaly detection, and demand forecasting. Additionally, FL ensures privacy preservation, captures local context, and promotes collaborative learning, making it a valuable tool for creating accurate predictive models in dynamic transportation environments.

\subsection{Scalability and Resource Efficiency}

FL is a data processing architecture that can be scaled up to train models on edge devices or nodes by utilizing distributed data. This approach allows for a decentralized system that reduces the need for data transmission to a central server, which in turn improves scalability and reduces network congestion. FL optimizes computing resource allocation by utilizing idle or underutilized resources on edge devices \cite{paper_013}. This efficient use of computing power minimizes infrastructure requirements and enhances resource management.

\section{Responsible AI Services for Smart Transportation}

\begin{figure}[t]
\centering
\includegraphics[scale=0.6]{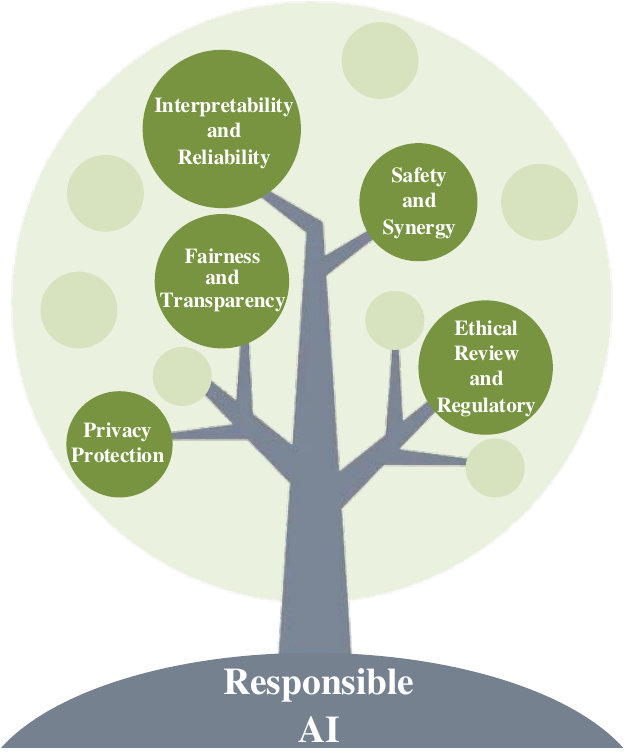} 
\caption{Advantages of responsible AI for smart transportation.}
\label{fig3}
\end{figure}

The application of responsible AI in intelligent transportation is expansive, striving to embed fairness, transparency, safety, reliability, and sustainability within decision-making processes and their outcomes, as shown in Fig. \ref{fig3}. Adherence to these principles within smart transportation systems fosters an improved transit ecosystem, ultimately advancing individual well-being.

\subsection{Privacy Protection Services}

In smart transportation, the extensive data collection, including vehicle location, speed, and trajectory, demands rigorous privacy and security protocols. Responsible AI services utilize methods like differential privacy and homomorphic encryption to address potential data breaches, misuse, and unauthorized access, thereby maintaining data confidentiality and integrity. For example, smart transportation systems process encrypted datasets, ensuring that sensitive information like location and speed remains encrypted even during intricate computations, minimizing the vulnerability to data leakage and unauthorized access.

\subsection{Fairness and Transparency Services}

The decision-making process in smart transportation should be transparent and unbiased to prevent discrimination and unjust outcomes. The system can use responsible AI services to ensure that its decision-making process and outcomes are impartial and transparent. This promotes collaboration and participation from multiple parties, leading to more open, fair, and sustainable decisions. For instance, when a traffic management system uses responsible AI services to optimize its traffic signal light scheduling, it employs a fair decision-making algorithm that considers the needs of all vehicles and pedestrians, avoiding any kind of discrimination.

\subsection{Interpretability and Reliability Services}

Responsible AI services can be integrated to achieve interpretability and reliability in smart transportation to ensure rational decision-making processes and transparent outcomes. This involves developing transparent decision models that consider various factors and implementing data preprocessing techniques for data consistency and accuracy, enhancing the reliability of the system. For example, by taking into account a variety of factors such as vehicle location, speed, driving rules, and environmental sensing data, smart transportation makes the decision-making process transparent so that individuals can understand the rationale behind the system's decisions, thereby improving interpretability.

\subsection{Safety and Synergy Services}

Smart transportation systems must prioritize safety and harmony to prevent traffic accidents and congestion. Responsible AI services can play a crucial role in improving the safety and harmony of the system by enabling information sharing and collaborative decision-making among vehicles and optimizing traffic signals. In a multi-vehicle cooperative driving system, responsible AI can significantly enhance collision avoidance and improve traffic flow by facilitating the necessary exchange of information among the vehicles. This exchange of information can enable better predictions and responses to each other's actions, leading to a safer and more efficient transportation system.

\subsection{Ethical Review and Regulatory Services}

The development and deployment of smart transportation need to be ethically reviewed and regulated to ensure that they meet ethical and legal standards. Responsible AI services can provide ethical review and regulation services, including assisting systems in developing ethical principles and guidelines, reviewing decision-making processes and outcomes, and more. For instance, responsible AI ensures the protection of individual privacy by monitoring the data collection and usage practices of the system, as well as conducting regular assessments and audits to ensure compliance with ethical and legal requirements.

\section{Responsible FL Service for Smart Transportation}

\begin{figure}[t]
\centering
\includegraphics[scale=0.6]{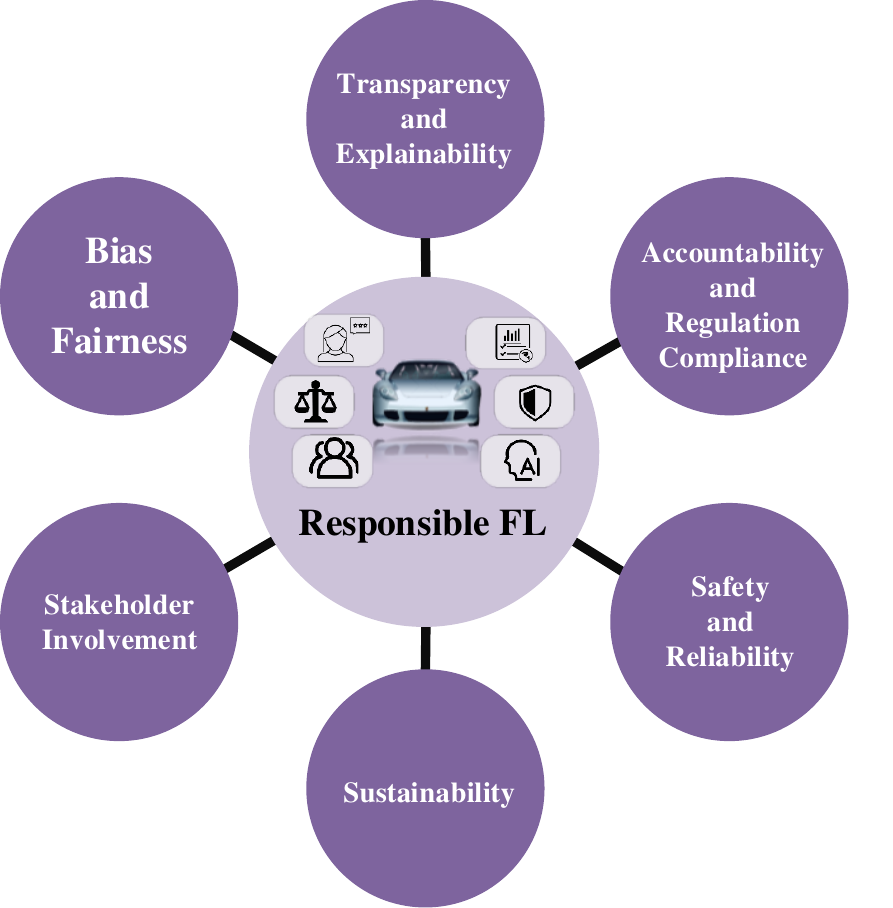} 
\caption{Advantages of responsible FL for smart transportation.}
\label{fig4}
\end{figure}

Integrating FL with Responsible AI is crucial in developing responsible FL for smart transportation systems. These systems handle sensitive data and have a direct impact on public safety and mobility. Responsible FL ensures that data used to optimize traffic, manage fleets, and inform travelers is processed while respecting privacy and minimizing biases. This is essential in a sector where equitable service provision is key. Moreover, transparent processes are necessary for users and regulators to understand and trust how data informs traffic patterns and decisions. Accountability is crucial when AI-made decisions have real-world consequences on traffic efficiency, emergency response times, and public transport reliability. In this section, we will discuss the key aspects of responsible FL in smart transportation systems, as shown in Fig. \ref{fig4}.

\subsection{Bias and Fairness}

The use of FL models to predict traffic patterns can be influenced by the diversity of data collected from different sources, such as vehicles and traffic sensors, which may have varying geographic, traffic, and demographic conditions. This diversity can lead to biases in the model, resulting in inaccurate traffic predictions for certain locations or demographic groups. To ensure fairness and accuracy in these models, it is important to adopt responsible AI practices that include implementing rigorous monitoring and rectification strategies \cite{paper_014}. These strategies may involve using advanced algorithms to identify and correct data disparities, such as reweighting data contributions from diverse sources or applying metric learning techniques to calibrate the model's predictive outputs.

\subsection{Transparency and Explainability}

The nature of training data in FL being distributed across various nodes can make it difficult to understand the decision-making process of the model. This can complicate the explanation of traffic decisions or route recommendations. To address this issue, responsible AI incorporates several approaches. Firstly, by using model interpretation techniques such as local feature importance analysis or decision tree explanations, the rationale behind the model's decisions can be elucidated. Secondly, increasing transparency through practices such as maintaining decision logs or creating detailed explanation reports offers users a clearer understanding of the model's outputs. These strategies aim to demystify the AI decision-making process, encouraging user trust and comprehension.

\subsection{Accountability and Regulation Compliance}

In smart transportation systems that use FL, accountability is crucial. It ensures that any system failures or misjudgments can be traced back to the responsible entities. Compliance with regulations is equally important as it helps to ensure adherence to traffic laws, safety standards, and data privacy mandates. Responsible FL should include mechanisms for legal compliance and accountability within its operational framework, establishing transparent processes that facilitate auditability and resolution in the event of discrepancies or incidents. This twin focus on accountability and regulatory compliance not only protects users and stakeholders but also reinforces the reliability and trustworthiness of intelligent transportation systems.

\subsection{Safety and Reliability}

Responsible FL is essential for ensuring individual and collective safety and reliability in smart transportation systems. At an individual level, it enables vehicles or sensors to operate with up-to-date and localized data, which enhances real-time decision-making for safety-critical applications. At the overseeing entity level, typically a central server, responsibility entails enforcing stringent standards for data integrity and model robustness. This dual focus ensures that each participant contributes to and benefits from a transportation network that learns and evolves without compromising the safety and privacy of individual data. Therefore, responsible FL not only reinforces the security of each contributing node but also upholds the trustworthiness and resilience of the collective transportation infrastructure.

\subsection{Sustainability}

Responsible FL is an approach that supports the sustainability of smart transportation systems. It achieves this by promoting efficient data usage and ensuring long-term operational viability. The approach involves processing data locally and updating models collectively. This way, it reduces the need for extensive data transmission, thus saving energy and network resources. By adopting this localized approach, the system extends the lifespan of devices through lower bandwidth consumption. It also promotes scalability, allowing the system to accommodate growing data volumes without incurring proportional increases in resource expenditure. Moreover, responsible FL practices ensure that smart transportation systems evolve within ethical and ecological constraints. This helps to minimize environmental impact while maximizing social benefit. Therefore, it is essential to develop sustainable smart transportation solutions equipped to meet future demands while preserving resources and prioritizing ethical considerations.

\subsection{Community Engagement and Stakeholder Involvement}

Effective FL in smart transportation systems requires the active participation of all parties including the government, private sectors, and individual users. It is important to follow responsible guidelines and regulations to ensure that everyone involved contributes in a way that is ethical, fair, and protective of privacy. By mandating compliance with responsible AI principles, responsible FL models establish a culture of accountability and shared ethical standards. This regulatory alignment ensures that collective efforts in model sharing, model training, and system utilization have a positive impact on the community, promoting trust and cooperation among all stakeholders in the smart transportation ecosystem.

\section{Challenges and Potential Soultions}

\begin{figure*}[t]
\centering
\includegraphics[scale=0.9]{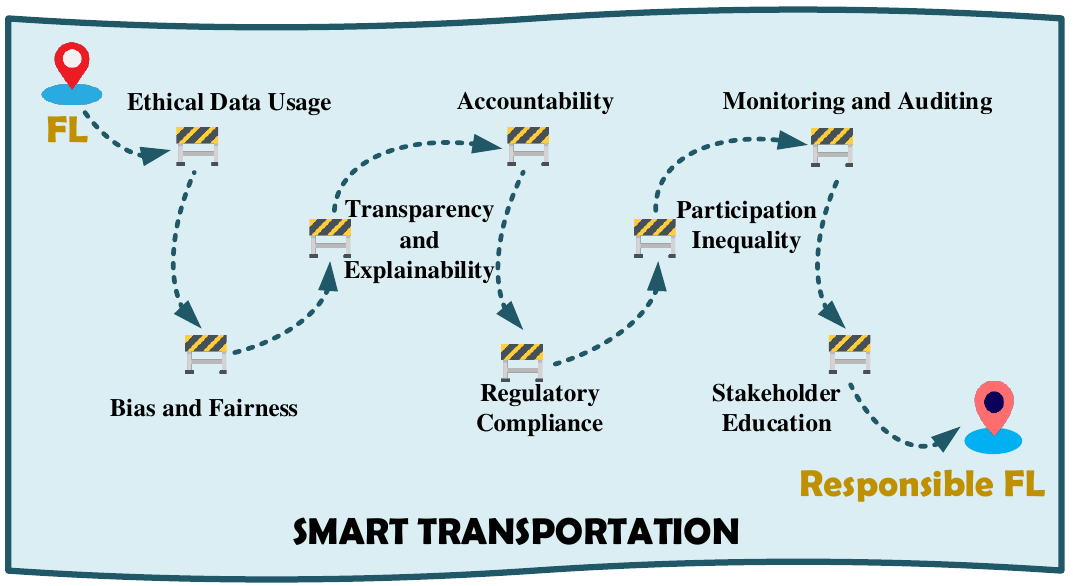} 
\caption{Journey from FL to responsible FL.}
\label{fig5}
\end{figure*}

Integrating responsible AI into FL within smart transportation systems presents a unique set of challenges that must be addressed to ensure ethical, fair, and effective use of technology. As smart transportation continues to evolve and become more data-driven and interconnected, the implementation of responsible AI principles in FL becomes increasingly crucial. However, this integration is not without its complexities, ranging from ethical data management to ensuring equitable system participation. Addressing these challenges is essential for the successful deployment of FL in smart transportation, where the stakes involve public safety, privacy, and sustainable urban development. The following points delve into these challenges in greater detail, as shown in Fig. \ref{fig5}, highlighting their significance and proposing potential solutions.

\subsubsection{Ethical Data Usage}

In smart transportation, data is sourced from a variety of nodes like vehicles, sensors, and public transit systems. This data is often collected under diverse conditions, leading to varying levels of consent and awareness among users \cite{paper_015}. The challenge arises in ensuring that this data is used ethically, respecting the privacy and consent of individuals. Ethical data usage is fundamental to building public trust and ensuring compliance with legal standards. To address this issue, it is crucial to implement a standardized ethical framework across the system. This framework should include clear protocols for obtaining consent and guidelines for ethical data processing. Doing so not only aligns with legal requirements but also enhances user trust in the system.

\subsubsection{Bias and Fairness}

It is essential to acknowledge that the diversity of data in FL systems, especially in smart transportation, can unintentionally create biases in AI models. This is because data from different geographic regions or demographic groups may not be equally represented. Such biases can lead to unfair or ineffective system performance, potentially disadvantaging certain user groups. Addressing bias is essential to ensure the service is equitable and maintain the system's credibility. Potential solutions to this problem include using advanced algorithms for bias detection and employing mitigation techniques such as adjusting the representation of underrepresented groups in the data. Ensuring fairness in AI models is not only an ethical responsibility, but also crucial for the long-term acceptance and effectiveness of the system.

\subsubsection{Transparency and Explainability}

The nature of FL models, particularly in the context of smart transportation, is decentralized and complex. This poses a significant challenge in terms of maintaining transparency and explainability. Both users and regulators may find it challenging to comprehend how AI makes decisions, which can create trust issues and hinder the acceptance of the system. To tackle this challenge, it is essential to ensure transparency and explainability by developing tools and standards that make the AI's decision-making process more interpretable and understandable. By enhancing the system's transparency, users will be able to trust and interact with the technology, which will lead to broader acceptance and more effective implementation.

\subsubsection{Accountability}

FL in smart transportation involves multiple contributors. However, it can be challenging to determine accountability when accidents or errors occur. To ensure legal and ethical compliance, it is essential to identify which participant or node is responsible for a particular decision or error. Maintaining the system's integrity and handling incidents effectively depends on clear accountability. To address this issue, it is necessary to establish well-defined legal and operational protocols that outline the responsibilities of all parties involved in the FL process. Such protocols ensure that liability is clear and a framework is in place to address errors effectively.

\subsubsection{Regulatory Compliance}

The use of FL in smart transportation is subject to a complex regulatory landscape at the international, national, and local levels that govern data protection and AI ethics. The main challenge lies in ensuring that the system complies with different laws that may vary across different regions. Achieving regulatory compliance is crucial to avoid legal issues and ensure that the system can be widely adopted. The solution is to create flexible and adaptable compliance frameworks that can dynamically adjust to different regulatory environments. This will ensure that the system remains compliant and functional across different jurisdictions.

\subsubsection{Participation Inequality}

In FL systems, there are often differences in the participation and influence levels among different stakeholders due to resource disparity. This can lead to an unequal distribution of power, where larger entities may dominate the system, potentially biasing it to their benefit. It is crucial to ensure fair participation to create a democratic and equitable system that serves the diverse needs of all users. To achieve this goal, it is vital to design incentive mechanisms that encourage and facilitate equal participation from all entities, regardless of their size or resources. This creates a more balanced system that accurately reflects and serves the interests of the entire community.

\subsubsection{Monitoring and Auditing}

Continuous monitoring and auditing of a decentralized FL system for adherence to responsible AI practices is a challenging task. Regular oversight is necessary to ensure that the system operates ethically and maintains its standards over time. The implementation of technologies such as blockchain can facilitate this by providing a transparent and verifiable record of the system's operations. Such technologies enable easier auditing and monitoring of the system, ensuring its compliance with responsible AI principles.

\subsubsection{Stakeholder Education}

Implementing responsible FL in smart transportation faces a significant challenge due to different levels of understanding about responsible AI principles among stakeholders. Educating all participants about these principles is crucial to ensure informed and ethical participation. Comprehensive educational programs and accessible resources can raise awareness and understanding of responsible AI practices. This will help to foster a culture of responsibility and ethics within the system, not only ensuring informed participation but also promoting ethical practices.

To implement responsible FL in smart transportation systems, it is crucial to address the above challenges. This includes ensuring ethical use of data, fairness, transparency, accountability, regulatory compliance, equal participation, continuous monitoring, and educating stakeholders. By doing so, the system can operate effectively and ethically, benefiting all its users. In the process of gradually realizing smart transportation, the aforementioned challenges do not arise at the outset or can be resolved simply once. For instance, in the pre-production phase, participants and programme designers need to sign relevant agreements. Based on the varying degrees of user consent in different data processing procedures and the difficulties in clearly defining responsibilities during accident scenarios, challenges related to ethical data usage (Challenge 1) and accountability (Challenge 4) may arise. During the algorithm design phase, the complex traffic situations and auditing processes impose high demands on distributed training processes (Challenge 3) and blockchain design (Challenge 7). In the initial design and later implementation phase, operators need to comply with the relevant regulations (Challenge 5),  encourage the involvement of all entities, and enhance stakeholders' understanding of the system (Challenge 6). Furthermore, challenges such as the improvement of regulations, system compatibility with diverse policies in different regions, and system fairness (Challenges 2, 5, 8) persist throughout the long-term operation of the system. Among them, once the regulations are updated (Challenge 5), Challenges 1, 4, and 7 may reemerge.

\section{Conclusion}

In conclusion, this paper has explored the integration of responsible AI and FL in smart transportation, highlighting the importance of their responsible use for system stability and sustainability. Although research in this domain is in the early stages, our study has critically examined the separate and combined impacts of FL and responsible AI on smart transportation. Key findings include improved fairness, transparency, interpretability, reliability, sustainability, accountability, regulation, and notably, enhanced community engagement and stakeholder involvement. We also identified ongoing challenges and proposed solutions, emphasizing that these challenges are not isolated incidents but continuous considerations in long-term system operations. Addressing and adapting to these challenges is essential for the continued success and effectiveness of smart transportation initiatives.


\section*{Biographies}

XIAOWEN HUANG (huangxiaowen@mail.dhu.edu.cn) is currently pursuing the Ph.D. degree in information and communication intelligence systems from Donghua University, Shanghai, China. Her research interests include edge computing, cognitive radio networks, and wireless communications.

TAO HUANG (tao.huang1@jcu.edu.au) is a Senior Lecturer at James Cook University, Cairns, Australia. He was an Endeavour Australia Cheung Kong Research Fellow, a visiting scholar at The Chinese University of Hong Kong, a research associate at the University of New South Wales, and a postdoctoral research fellow at James Cook University. Dr. Huang is an Associate Editor of the IEEE Open Journal of Communications Society, IEEE Access, and IET Communications.

SHUSHI GU (gushushi@hit.edu.cn) is an Associate Professor with the School of Electronic and Information Engineering in Harbin Institute of Technology (Shenzhen), Shenzhen, China, and the researcher with Guangdong Provincial Key Laboratory of Aerospace Communication and Networking Technology, Shenzhen, China. From 2018 to 2019, he was a Postdoctoral Research Fellow with James Cook University, Cairns, Australia. His current research interests include satellite communications, network coding theory, satellite-terrestrial integrated network, and distribute storage and computing.

SHUGUANG ZHAO (sgzhao@dhu.edu.cn) is currently a Professor with the College of Information Science and Technology, Donghua University, Shanghai, China. His research interests include electronic design automation (EDA$\&$ESL), programmable device and embedded system development, intelligent information processing, pattern recognition and intelligent system.

GUANGLIN ZHANG (glzhang@dhu.edu.cn) is currently a Professor and the Vice Dean with the College of Information Science and Technology, Donghua University, Shanghai. His research interests include online algorithms, capacity scaling of wireless networks, vehicular networks, smart microgrids, and mobile edge computing. He has been the Local Arrangement Co-Chair of ACM TURC 2017 and 2019 and the Vice Technical Program Committees Co-Chair of ACM TURC 2018 and 2021. He is an Editor on the Editorial Board of IEEE/CIC CHINA COMMUNICATIONS.

\end{document}